\documentclass[11pt]{article}
\usepackage[margin=1in]{geometry}
\usepackage{amsmath,amssymb}
\usepackage{graphicx}
\usepackage{cite}
\usepackage{hyperref}
\usepackage{pgfplots}
\usepackage{subcaption}
\pgfplotsset{compat=1.17}
\usepackage{graphicx}
\usepackage{listings}
\title{\textbf{Deterministic and Stochastic Models in Enzyme Kinetics}\\
An Expanded Analysis with Simulation and Experimental Context}

\author{Jiaji Qu and Malini Rajbhandari}
\date{March 1, 2025}

\begin{document}

\maketitle

\begin{abstract}
Enzyme kinetics has historically been described by deterministic models, with the Michaelis--Menten (MM) equation serving as a paradigm. However, recent experimental and theoretical advances have made it clear that stochastic fluctuations, particularly at low copy numbers or single-enzyme levels, can profoundly impact reaction outcomes. In this paper, we present a comprehensive view of enzyme kinetics from both deterministic and stochastic perspectives. We begin by deriving the classical Michaelis--Menten equation under the quasi-steady-state assumption (QSSA) and discuss its validity. We then formulate the corresponding stochastic model via the chemical master equation (CME) and illustrate how the Gillespie algorithm can simulate single-molecule events and briefly use Kampen's system-size expansion to justify our simulation methods. Through extended computational analyses---including variance calculations, phase-plane exploration, and parameter sensitivity---we highlight how deterministic and stochastic predictions coincide in certain limits but can diverge in small systems. We further incorporate case studies from single-enzyme turnover experiments and cellular contexts to showcase the real-world implications of noise. Taken together, our results underscore the necessity of a multifaceted modeling strategy, whereby one can switch between deterministic methods and stochastic realism to gain a fuller understanding of enzyme kinetics at different scales.
\end{abstract}

\pagebreak

\tableofcontents

\pagebreak

\section{Introduction}

\hspace{0.5in} An enzyme is a biological unit that catalyzes biological reactions. It does so by lowering the activation
energy that is required for the reaction to occur. Thus, with the help of enzymes, reactions are able to 
occur much faster than they would otherwise. The study of mechanics of these enzymes, which includes
how fast the enzyme catalyzes reactions under different conditions is known as enzyme kinetics. Briggs and Haldane (1925) were the first people to make the connection between these enzyme reactions on the chemical level and the dynamics of these enzyzmes on a biological level and result in the notation we use in section 3.2. 

\section{Problem Formulation and Models}
\hspace{0.5in} Traditionally, the Michaelis–Menten (MM) model is used to describe how the rate of product formation 
depends on substrate concentration under conditions where enzyme is limiting. Deterministic models
based on ordinary differential equations (ODEs) which are able to capture average behavior in large, 
well-mixed volumes, leading to widespread use of the MM equation in a vast number of biologically 
relevant fields. 

However, recently, research on single-molecule experiments have revealed that enzymes can exhibit 
significant fluctuations at the molecular scale. In many cellular contexts, enzyme or substrate numbers may
be small (tens or hundreds of molecules). This makes the molecular noise in these contexts non-negligible. 
So, in many cases, the deterministic models are not as impactful in capturing the whole picture regarding 
the enzyme kinetics. Stochastic equations fill that void. They typically employ the chemical master equation
(CME) appearing in the form of stochastic simulation algorithms become essential for describing such 
situations. In this paper, we present a comparison and review of deterministic and stochastic models in enzyme
kinetics, emphasizing mathematical derivations and experimental relevance. 
\hspace{0.5in} 

\subsection*{Models (type A and B)}

\textbf{Type A} models, also known as the Michaelis Meneten equation (with differing coefficients), were initially derived in the seminal paper "Die Kinetik der Invertinwirkung." by Michaelis, L., \& Menten, M. (1913). These models have been adapted and adjusted as time has passed, we will explore the transformation of this model type though these years. We begin by recalling the mass-action ODE formulation
for a simple enzyme–substrate reaction and derive the Michaelis–Menten rate law via the quasi-steady-state 
assumption to give a solution to the model. 

Next, we then introduce the stochastic viewpoint, formulating \textbf{Model type B} which is known as the Chemical Master Equation (CME) and discussing the Gillespie algorithm as a Monte Carlo method for simulating exact stochastic trajectories. We will briefly touch on the ideas behind the the proof of the CME but more importantly we will give a full proof as to  why Monte-Carlo methods work under large enough sample size and discuss how to quantify fluctuations (variance, waiting times, etc.)and compare deterministic and stochastic dynamics through computational examples, phase-plane analyses, and parameter sensitivity. 

Finally, we incorporate 
real experimental data  (e.g., single-enzyme turnover) to illustrate how molecular noise shows up in practice and how
it can still be consistent with Michaelis–Menten on average. Our aim is to highlight both the power and limits of 
deterministic enzyme kinetics, and to demonstrate the complementary role of stochastic modeling in capturing 
small-scale noise-driven phenomena.

\section{Solution to Model A: Deterministic Enzyme Kinetics}
\subsection{Enzyme Catalyzed Reaction}
Enzyme catalyzed reactions, as described above, are everywhere in biochemistry. An enzyme-substrate reaction set is used to model this system:\\
\begin{equation}
E + S \;\underset{k_{-1}}{\stackrel{k_1}{\rightleftharpoons}}\; ES 
\;\xrightarrow{k_2}\; E + P,
\end{equation}
where $E$ is the free enzyme, $S$ is the substrate, $ES$ is the enzyme--substrate complex, and $P$ is the product. The rate constants $k_1$ (binding), $k_{-1}$ (unbinding), and $k_2$ (catalytic). The net result of this reaction is the conversion of substrate to product. Note that the enzyme is unconsumed and can be recycled in future reactions (Haas).\\
There are many factors that can impact enzyme activity. Each reaction has an optimal temperature and pH range. For instance high temperatures usually help to speed up reactions but if they are too high, this may cause the enzyme to denature and stop working all together. Similarly with pH, if an enzyme is in conditions outside its' natural pH rate, it will breakdown and not function properly. This is also true for other circumstances like concentration of the enzymes or the substrate concentration.\\

\subsection{Mass-Action ODE Formulation}
Before deriving the differential equations, let us briefly introduce the law of mass action. This law states that the rate of a reaction is proportional to the product of the concentrations of the reactants. This principle allows us to model enzyme kinetics mathematically.\\
Let us now look back at equation 1:
\begin{equation}
E + S \;\underset{k_{-1}}{\stackrel{k_1}{\rightleftharpoons}}\; ES 
\;\xrightarrow{k_2}\; E + P,
\label{eq:basic_reaction}
\end{equation}
 We will now focus on the fact that the rate constants $k_1$ (binding), $k_{-1}$ (unbinding), and $k_2$ (catalytic) follow the law of mass action in deterministic models. This means that one is able to derive the ODE equations for the given equation by using the given equation. So, for equation 1, let $[S](t), [ES](t), [P](t)$, and $[E](t)$ denote concentrations at time $t$. \\
 Then the ODEs are:
\begin{align}
\frac{d[S]}{dt} &= -k_1 [E][S] + k_{-1}[ES], \label{eq:dSdt}\\
\frac{d[ES]}{dt} &= k_1 [E][S] - (k_{-1} + k_2)[ES], \label{eq:dESdt}\\
\frac{d[P]}{dt} &= k_2 [ES],\label{eq:dPdt}
\end{align}
\begin{itemize}
    \item Equation (3) describes how the substrate concentration decreases as a result of binding with the enzyme and increases as a result of unbinding of the enzyme-substrate complex.
    \item Equation (4) represents the formation of the enzyme-substrate complex and its subsequent breakdown either back to free enzyme and substrate or forward to product formation.
    \item Equation (5) tracks the rate at which the product is formed from the enzyme-substrate complex.
\end{itemize}
Since the total enzyme concentration is conserved, we express free enzyme concentration as:\\
 $$[E] = [E]_T - [ES]$$ 
 We can see that the rate constants become the coefficients for the ODE and the reactants and products become the concentrations. Thus, these ODEs are able to give us the rate of change in concentration of each element of this reaction. This is key in being able to track the progress of a reaction. The equation allows for further analysis, such as deriving the Michaelis-Menten equation under steady-state assumptions.\\

\subsection{Quasi-Steady-State Assumption and the Michaelis--Menten Equation}
The Michelis-Menten model is a very well-known and simple approach to enzyme kinetics. It looks at reaction velocity and substrate concentration, focusing on the relationship between them. The Michaelis-Menten equation for the system given by equation 1 is:
\[
v \;=\; \frac{V_{\max} [S]}{K_M + [S]}.
\]
We will now be discussing the George Briggs and J.B.S. Haldane derivation of this equatuion from 1925. This equation utilizes what is known as the approximation. This is the assumption that the concentration of the enzyme-substrate complex $[ES]$ will rapidly approach a steady state as a significant amount of substrate is consumed, after which the rate of change is assumed to be 0. So:
$$d[ES]/dt \approx 0$$ 
This then implies that:
$$k_{1} [E][S] = \left(k_{-1} + k_{2}\right)[ES]$$

We will start with this assumption for the derivation and use basic algebra to rearrange and solve for $v$. Let us now look at the derivation (Atkins):
\begin{align*}
    k_1 ([E_T] - [ES])[S] &= (k_{-1} + k_2)[ES] \\
    k_1 [E_T][S] - k_1 [ES][S] &= (k_{-1} + k_2)[ES] \\
    k_1 [E_T][S] &\approx (k_{-1} + k_2)[ES] + k_1 [ES][S] \\
    [ES] &= \frac{k_1 [E_T][S]}{(k_{-1} + k_2) + k_1 [S]} \\
    &= \frac{[E_T][S]}{\left(\frac{k_{-1} + k_2}{k_1}\right) + [S]} \\
    \Rightarrow v &= k_2 [ES] = \frac{k_2 [E_T][S]}{\left(\frac{k_{-1} + k_2}{k_1}\right) + [S]}
\end{align*}\\

\pagebreak
Once we have the equation for $v$, we are able to substitute in $[E]=[E]_T - [ES]$ and use that to solve for $[ES]$:
\[
[ES] \;\approx\; \frac{[E]_T [S]}{\frac{k_{-1}+k_2}{k_1} + [S]} = \frac{[E]_T [S]}{K_M + [S]},
\]
Then, we use that $K_M \equiv (k_{-1}+k_2)/k_1$ is the Michaelis constant. The rate of product formation is then given by
\[
v = \frac{d[P]}{dt} = k_2 [ES] = k_2\, \frac{[E]_T [S]}{K_M + [S]}.
\]
Defining $V_{\max} = k_2 [E]_T$, the familiar Michaelis--Menten equation follows:
\[
v \;=\; \frac{V_{\max} [S]}{K_M + [S]}.
\]
This derivation relies on two assumption. The first is the quasi-steady-state assumption that $\frac{s[ES]}{dt} = 0$, typically valid when $[E]_T \ll [S]$ or when binding/unbinding is fast compared to the catalytic turnover. Additionally, a second assumption that is important to this derivation is that $[S]$ represents the free substrate concentration which for this case is assumed to be close to the total substrate concentration present. This is called the free ligand approximation, and is valid as long the total enzyme concentration is well below the $K_M$ of the system. \\
When talking about the Michaelis-Menten equation, it is imperative to also discuss the graph that arises from this model known as the Michaelis-Menten Curve. This graph helps contextualize the variables in the equation above.
\begin{center}
   \includegraphics[scale = 0.3]{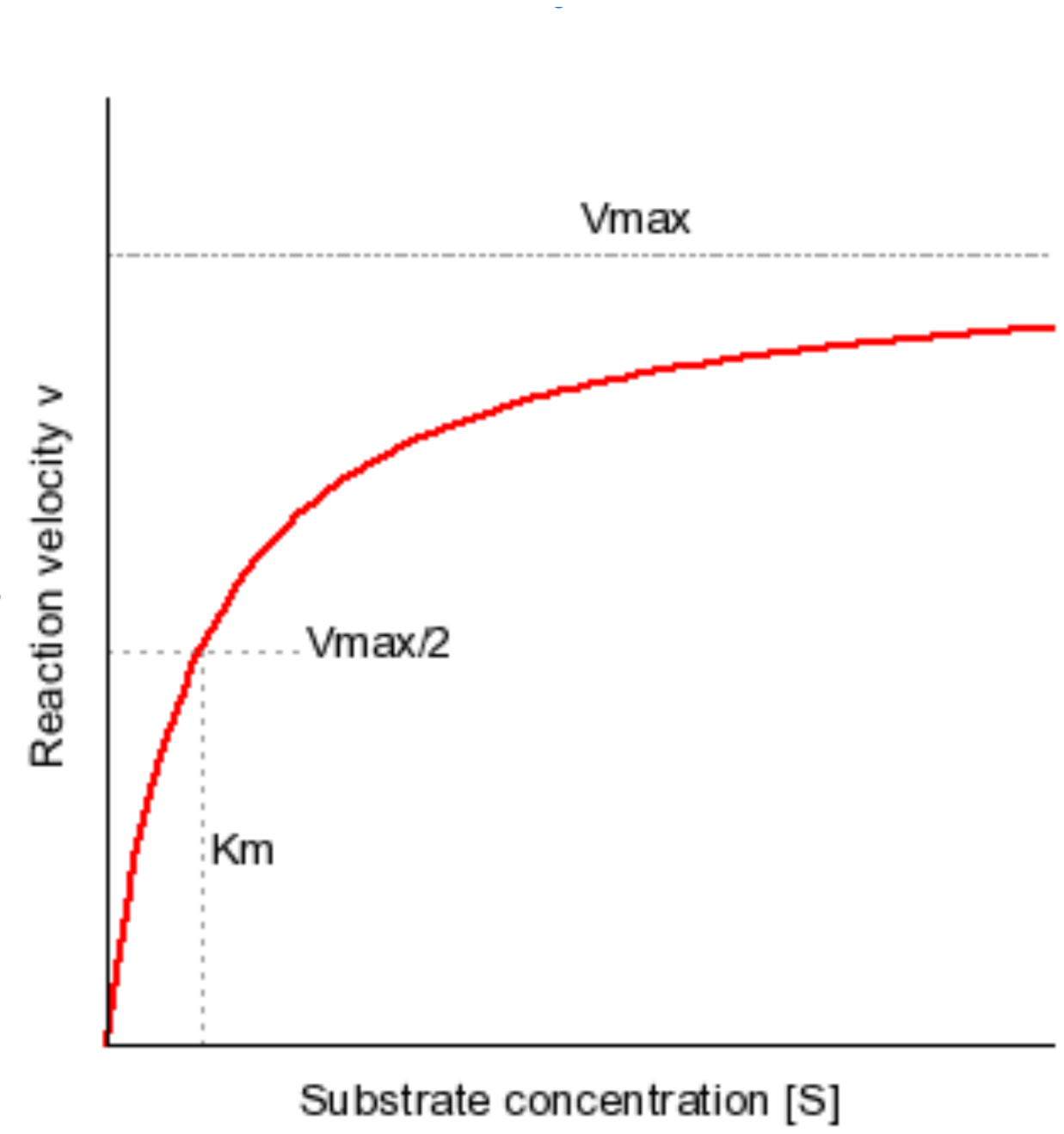}
\end{center}
Here, we have a lot of the variables used above visualized on a graph. We see that the reaction velocity, $v$ is the y-axis and the substrate concentration $[S]$ is the x-axis. Vmax is the maximum velocity achieved by the system. Finally, $K_M$ is the substrate concentration at which half of the Vmax is achieved.

\section{Solution to Model B: Stochastic Enzyme Kinetics: Chemical Master Equation and SSA}
\subsection{Motivation for Stochastic Models}
Inside living cells, volume scales are extremely small, and copy numbers of enzymes or substrates may be in the tens to hundreds. Random collisions and reaction events are inevitable and cause \emph{intrinsic noise}. Deterministic ODEs such as Michaelis-Menten track average behavior but does not account for the random noise and thus cannot capture the distribution of outcomes or the probability of rare events. Stochastic models, formulated via the chemical master equation (CME), treat reaction events probabilistically and track the time evolution of $P(n_E,n_S,n_{ES},n_P;t)$, the probability of having integer counts of each species.

\subsection{Chemical Master Equation (CME) for Enzyme Kinetics}
For the mechanism in Eq.~\eqref{eq:basic_reaction}, define $n_S, n_E, n_{ES}, n_P$ as the molecule counts. The propensity functions (reaction rates) are (Grimma, 2009)) :
\[
\begin{aligned}
&a_1 = k_1\,n_E\,n_S \quad (\text{for } E + S \to ES),\\
&a_2 = k_{-1}\,n_{ES} \quad (\text{for } ES \to E + S),\\
&a_3 = k_2\,n_{ES} \quad (\text{for } ES \to E + P).
\end{aligned}
\]
The CME is an infinite-dimensional system of ODEs for $P(n_E,n_S,n_{ES},n_P;t)$:
\[
\frac{d}{dt}P(\mathbf{n};t) \;=\; \sum_{r} \Big[\, W_{r}(\mathbf{n}-\boldsymbol{\nu}_r)\,P(\mathbf{n}-\boldsymbol{\nu}_r;t) \;-\; W_{r}(\mathbf{n})\,P(\mathbf{n};t)\Big],
\]
where $\mathbf{n}=(n_E,n_S,n_{ES},n_P)$, $\boldsymbol{\nu}_r$ is the stoichiometric change due to reaction $r$, and $W_r(\cdot)$ are the propensities $a_r$. This fully characterizes the probabilities of all states over time. Solving the CME analytically is generally impossible for large systems, but it remains the gold-standard microscopic description (Gillespie, 2000).

\subsection{Gillespie's Stochastic Simulation Algorithm (SSA)}
In practice, one often uses \emph{stochastic simulation} rather than solving the CME directly. Informally, Gillespie's SSA (Gillespie, 1997) generates simulated trajectories by:
\begin{enumerate}
\item Calculating propensities $a_r$ in the current state.
\item Sampling the waiting time to the next reaction from an exponential distribution with mean $1/a_0$, where $a_0=\sum_r a_r$.
\item Randomly selecting which reaction occurs, weighted by $\frac{a_r}{a_0}$.
\item Updating the state and repeating until a final time is reached.
\end{enumerate}
The SSA yields a piecewise-constant trajectory for $(n_E(t),n_S(t),n_{ES}(t),n_P(t))$. By averaging over many runs, we can approximate means and variances of species counts (as a result, this is a form of \textbf{Monte Carlo} simulation). Moreover, as molecule numbers grow large, results converge to the deterministic ODE solution (law of large numbers). At small copy numbers, however, large stochastic fluctuations emerge, causing outcomes to diverge run-to-run. 

In this section we shall provide a derivation of Gillespie's SSA and some examples in how it's used compared to the ODEs method. 

\subsubsection{Derivation}

The CME describes how the probability $P(\mathbf{n},t)$ of being in a particular state $\mathbf{n}$ (specifying the molecule counts of each species) evolves over time. For a well-mixed volume, each reaction channel $r$ has a \emph{propensity} function $a_r(\mathbf{n})$ which gives the probability per unit time of that reaction firing when the system is in state $\mathbf{n}$. Summing over all reactions gives the total propensity
\[
a_0(\mathbf{n}) \;=\; \sum_{r=1}^{R} a_r(\mathbf{n}).
\]
The \emph{key insight} is that in a Markov process with total rate $a_0$, the waiting time to the next reaction event follows an exponential distribution with mean $1/a_0$. Moreover, the probability that the next reaction to occur is of type $r$ is $a_r / a_0$. Gillespie's SSA implements this logic directly, thus providing a \emph{direct Monte Carlo} simulation of the exact solution to the CME.

\subsubsection{Naive (direct SSA) (``Gillespie'') Algorithm}

\begin{enumerate}
    \item \textbf{Initialize.} Set the initial time $t = t_0$ and the initial state $\mathbf{n}(t_0)$ (the molecule counts of each species). Determine the maximum simulation time $t_{\mathrm{end}}$ or stopping criterion (e.g., depletion of a key substrate).
    \item \textbf{Calculate current propensities.} Evaluate $a_r(\mathbf{n}(t))$ for each reaction channel $r = 1,\dots,R$. Let
    \[
    a_0 = \sum_{r=1}^{R} a_r(\mathbf{n}(t)).
    \]
    \item \textbf{Sample next reaction time.} Draw a random number $r_1$ uniformly in $(0,1)$ and compute the time to the next reaction:
    \[
    \tau \;=\; \frac{1}{a_0}\,\ln\!\bigl(\frac{1}{r_1}\bigr) \;=\; -\frac{\ln(r_1)}{a_0}.
    \]
    This $\tau$ is exponentially distributed with mean $1/a_0$.
    \item \textbf{Sample which reaction fires.} Draw another random number $r_2$ uniformly in $(0,1)$. Determine the reaction channel $j$ such that
    \[
    \sum_{r=1}^{j-1} a_r < r_2 \,a_0 \;\le\; \sum_{r=1}^{j} a_r.
    \]
    This effectively selects a reaction $j$ with probability $a_j / a_0$.
    \item \textbf{Update state and time.} Advance the simulation time by $\tau$, i.e.\ $t \leftarrow t + \tau$, and update the state $\mathbf{n} \leftarrow \mathbf{n} + \boldsymbol{\nu}_j$, where $\boldsymbol{\nu}_j$ is the stoichiometric change vector for reaction $j$.
    \item \textbf{Repeat.} If $t < t_{\mathrm{end}}$ (and any other termination condition has not been met), return to Step~2. Otherwise, stop.
\end{enumerate}

This procedure yields a \emph{piecewise constant} trajectory for $\mathbf{n}(t)$, with jumps at each reaction event. Repeated runs from the same initial condition produce an \emph{ensemble} of possible trajectories, from which we can estimate means, variances, and other statistics.

\subsection{Some brief analysis of SSA}

The direct SSA is \emph{exact} in the sense that it correctly samples from the probability distribution governed by the CME, given the assumptions of well-mixedness and Markovian reaction events. However, if $a_0$ is large (many reactions firing in short intervals), a large number of events must be simulated, potentially making the method computationally expensive. Variants such as the \emph{first-reaction method}, \emph{next-reaction method}, and \emph{partial-propensity methods} exist to improve computational efficiency or to handle special cases (e.g., many reaction channels but relatively few actual firings).

\subsubsection{Alternative formulations of SSA}

If the reader has time, they may consider to implement a more sophisticated SSA for their own purposes. The scope of this paper merely implements the direct (naive) Gillespie. Here we list some other algorithms one can try.  

\paragraph{First-Reaction Method.}  
In the \emph{first-reaction method}, one conceptually draws a separate exponential waiting time for each reaction channel $r$, i.e.\ $\tau_r = -\ln(\rho_r)/a_r$, where $\rho_r$ is a random number in $(0,1)$. The channel $j$ with the smallest $\tau_j$ is the next reaction to occur, and the system updates accordingly. Mathematically, this yields the same distribution of next-event times and reaction channels as the direct method, though the implementation details differ.

\paragraph{Next-Reaction Method.}  
The \emph{next-reaction method} (Gibson and Bruck) reduces the computational cost of redrawing times for each reaction after every event by keeping a priority queue or other data structure. After a reaction fires, it updates only the affected reaction channels. This can offer efficiency gains in large networks but is more complex to implement.

\paragraph{Partial-Propensity Formulations.}  
For reactions that can be factored into sub-steps (especially in gene-regulatory networks), \emph{partial-propensity} approaches can further reduce computations. They separate each channel's propensity into smaller building blocks, updating only those parts affected by a state change. Wilkinson (2011) makes a mention of this briefly under his section on Hybrid methods (Chapter 8.4). 

\subsubsection{Chemical Master Equation (Large Sample Proof via van Kampen system-size expansion)}

Because the SSA samples exactly from the CME, it reproduces the \emph{master equation} solution in the limit of many trajectories. As molecule counts become large, one can show (via the law of large numbers) that the ensemble-averaged trajectory converges to the solution of the deterministic rate equations. The \emph{system-size expansion} (van Kampen) formalizes this connection, showing how the leading-order term recovers the ODEs, while next-order corrections yield Gaussian fluctuations of order $\Omega^{-1/2}$ (where $\Omega$ is system volume).

\section{Computational Experiments}
We now illustrate deterministic and stochastic enzyme kinetics via numerical experiments, focusing on variance calculations, bifurcation insights due to additional feedbak, and how parameter changes in $k_1, k_2, k_{-1}, \textbf{$[E]_T$}$ affect noise levels. We will be numerically integrating the ODEs in Eqs.~\eqref{eq:dSdt}-\eqref{eq:dPdt} which will give a smooth trajectory for $[S](t)$ and $[P](t)$. Meanwhile, we shall run N = 1000 Monte-Carlo Gillespie simulations to produce a distribution of $(n_S(t), n_P(t))$. The code for both the deterministic and stochastic methods will be in the Appendix. All of these stochastic algorithms come directly from Wilkinson's book.

\subsection{Deterministic ODE vs Gillespie SSA Example}
\label{sec:ode-ssa-example}
For (A) of Figure (1), let $[E]_T=1$~$\mu$M, $[S]_0=10$~$\mu$M in a volume that results in about 10--100 total molecules. For (B), say $[E]_T=100$~$\mu$M, $[S]_0=1000$~$\mu$M. Suppose $k_1=0.1,\;k_{-1}=0.05,\;k_2=0.05$ (arbitrary units). 

\begin{figure}[h]
\centering
\begin{subfigure}[b]{0.73\textwidth}
    \centering
    \includegraphics[width=\textwidth]{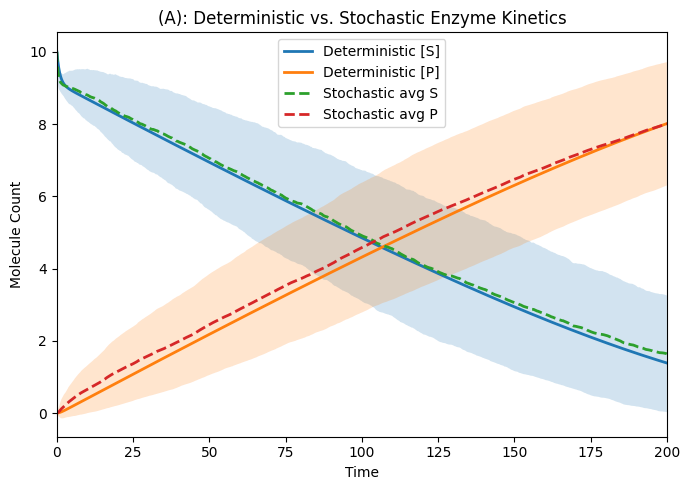}
    \caption{Low copy number}
    \label{fig:low_count}
\end{subfigure}
\end{figure}

\newpage

\begin{figure}[h]
\centering
\begin{subfigure}[b]{0.73\textwidth}
    \centering
    \includegraphics[width=\textwidth]{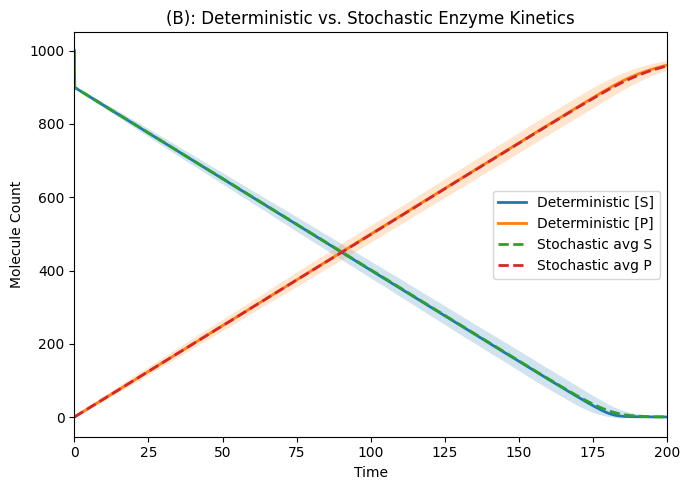}
    \caption{High copy number}
    \label{fig:high_copy}
\end{subfigure}7
\centering
\caption{Monte-Carlo simulation of stochastic vs deterministic enzyme kinetics for small vs large molecule numbers. 
Panel (A) shows a wide spread in $n_P(t)$, while Panel (B) shows relatively small fluctuations.}
\label{fig:small_vs_large}
\end{figure}

\newpage

\subsection{Variance and Probability Distributions}
Beyond mean trajectories, the variance $\mathrm{Var}[n_P(t)]$ quantifies the spread in product counts. Deterministically, there's no variance at all for a given initial condition. Stochastically, we observe:

\[
\mathrm{Var}[n_P(t)] \;=\; \langle (n_P(t) - \langle n_P(t)\rangle)^2\rangle.
\]

For a closed system, product eventually saturates at $n_{P,\text{max}} = n_{S0}$ (assuming complete conversion). The variance typically increases up to a certain point and then decreases as all substrate is consumed. For open systems (continuous substrate inflow), a nonzero steady-state variance emerges. System-size expansion or moment-closure can predict approximate analytic forms for these variances. Below is a an example of this coming from the same data as the previous simulation results. 

\begin{figure}[h]
\centering
\begin{subfigure}[b]{0.75\textwidth}
    \centering
    \includegraphics[width=\textwidth]{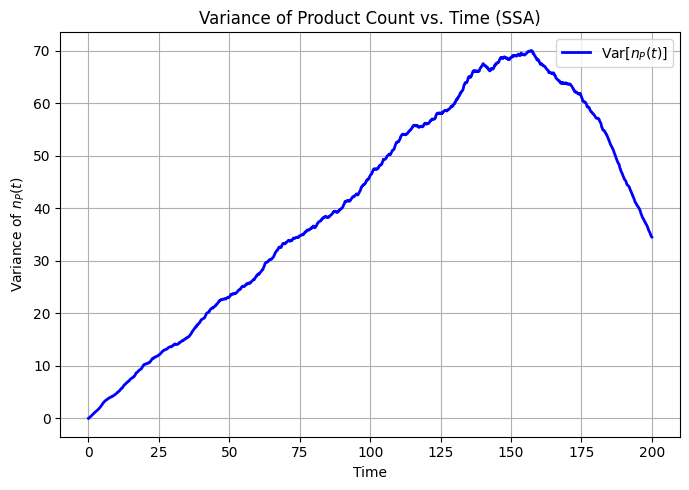}
    \caption{Time vs. Variance of Product Count}
    \label{fig:low_count}
\end{subfigure}
\end{figure}

\subsection{Phase-Plane and Bifurcation Considerations}
\label{sec:phaseplane}
For the basic mechanism \eqref{eq:basic_reaction}, the deterministic ODE typically has a single stable steady state (complete consumption if closed). Phase-plane analysis in $(S,ES)$ space shows a rapid transient toward the quasi-steady-state manifold $ES \approx \frac{[E]_T S}{K_M + S}$. More complex enzyme networks (e.g., with feedback or multiple substrates) can exhibit bistability or oscillations. Deterministically, these appear as bifurcations in the ODE system. Stochastic modeling then predicts random switching between stable states or noisy oscillations. While standard MM kinetics is too simple for such dynamics, advanced kinetic schemes do indeed show such emergent complexity. 

\vspace{0.15cm}

Below are two examples of this behavior where the first is a Phase Plane for a Simple Enzyme Kinetics described by an ODE system and the other is described by a random switching scheme.  

\begin{figure}[h]
    \centering
    \includegraphics[width=0.75\linewidth]{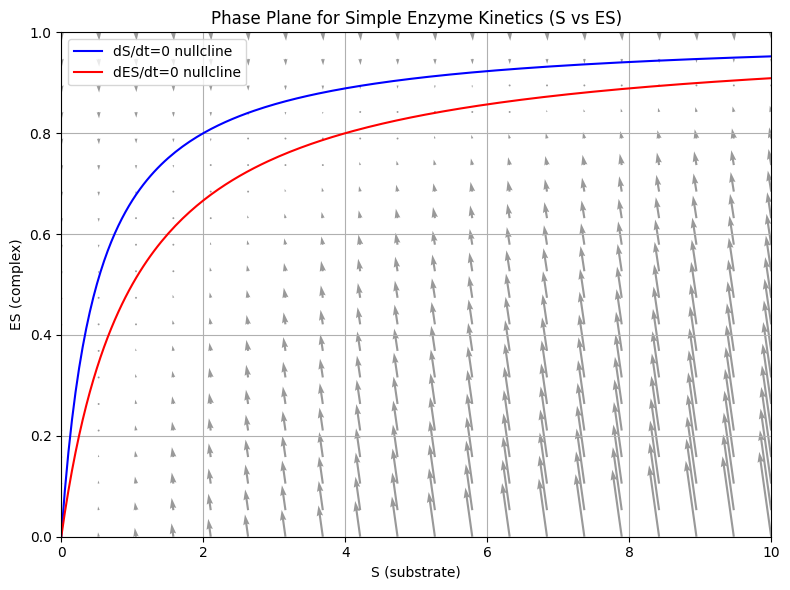}
    \label{fig:enter-label}
\end{figure}

\begin{figure}[h]
    \centering
    \includegraphics[width=0.90\linewidth]{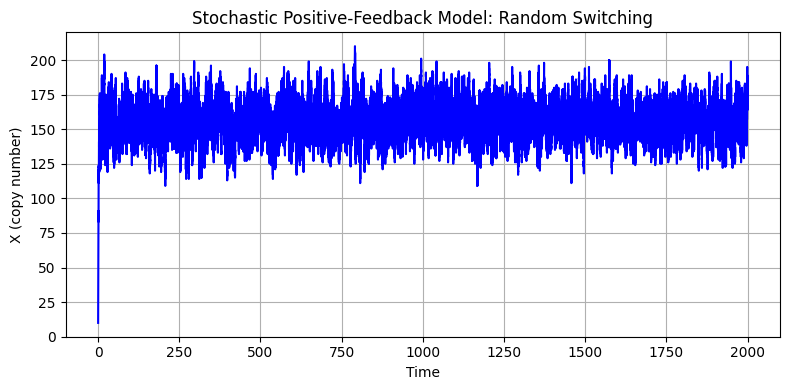}
    \label{fig:enter-label}
\end{figure}

\subsection{Parameter Sensitivity}
Deterministically, the time to consume half the substrate (or other metrics) can be analytically related to $K_M, V_{\max}$, etc. In the stochastic model, the distribution of reaction times depends on those same parameters, but also the \emph{initial molecule counts} and volume. Under low copy conditions, adjusting $k_2$ can have a disproportionate effect on the variance in product, not just the mean. \textbf{System-size expansion} clarifies that the magnitude of fluctuations scales as $\Omega^{-1/2}$ when the system is monostable, meaning that smaller volumes (or smaller total molecule counts) amplify noise. Real cells, however, employ regulatory strategies to mitigate large swings in enzyme expression. For the sake of length, a longer discussion of parameter sensitivity and sensitivity analysis can be found in van Kampen's book "Stochastic Processes in Physics and Chemistry" (2007) and Wilkinson (2011). 

\subsection{Algorithmic Complexity and Extensions}

Briefly, we shall discuss some issues related to the SS Algorithms themselves, namely related to their complexity and speed as well as overall stability. For more information, please see Wilkinson's book "Stochastic Modelling for Biology" published in 2011. 

\paragraph{Complexity.}  
Each iteration of the direct SSA takes $\mathcal{O}(R)$ operations to compute or update the sum of propensities. If many reactions fire in a short physical time, the total number of iterations can be large, leading to high computational cost for large systems or long timescales.

\paragraph{Approximate Methods (Tau-Leaping).}  
For systems where propensity functions change slowly relative to reaction firing rates, \emph{tau-leaping} is an approximate method that leaps forward in time by a fixed $\Delta t$ (``tau''), firing multiple reactions in bulk. This can drastically speed up simulations, though at the cost of exactness.

\paragraph{Hybrid Deterministic--Stochastic Schemes.}  
In biochemical networks where some species are abundant (and can be treated deterministically) while others are rare (requiring stochastic treatment), one can combine ODEs for the high-copy species with SSA for the low-copy species. Such hybrid approaches aim to balance accuracy with computational feasibility.

\section{Real Experimental Data}
\subsection{Single-Enzyme Turnover Studies}
Techniques such as fluorescence resonance energy transfer (FRET), optical trapping, and single-molecule fluorescence bursts have enabled direct observation of individual enzyme turnovers. For instance, English B.P. \emph{et al.}\ reported real-time records of $\beta$-galactosidase converting single substrate molecules (English, B.P. et al, 2006). The waiting times between product formation events were random, yet the \emph{mean} rate conformed to the Michaelis--Menten formula. In some cases, multi-exponential waiting time distributions revealed multiple rate-limiting steps or enzyme conformations (i.e., dynamic disorder). Deterministic ODEs cannot describe such discrete fluctuations, but the CME or extended stochastic models with additional enzyme states do.

\subsection{Noise in Cellular Metabolism}
In vivo, enzymes are part of larger metabolic pathways. Gene expression noise leads to cell-to-cell variation in enzyme abundance, substrate availability, etc. Observations of single-cell metabolite levels often show broad distributions. Stochastic enzyme models explain how random timing of enzyme-substrate encounters produces variability in product formation. Interestingly, \emph{averaged} data across cell populations may still fit an MM-type curve, while single-cell data exhibit significant scatter (Grimma, 2009). This phenomenon of population averaging is analogous to how the mean of many SSA trajectories matches the deterministic solution, but individual realizations vary.

\subsection{Integrated View: Michaelis--Menten on Average, Stochastic in Detail}
Across numerous studies, a consistent theme emerges: \emph{Michaelis--Menten remains valid for mean rates} over many molecules or many turnover events, but \emph{stochastic fluctuations} give additional insight into the timing and distribution of reaction events. Even single-enzyme data often yields an average turnover frequency consistent with $v = \frac{k_2[S]}{K_M + [S]}$ (English, B.P. et al., 2006). The distribution around this mean, however, may reveal deeper mechanisms such as hidden conformational states, partial reversibility, etc.. Thus, the deterministic framework suffices for many bulk kinetic analyses, while the stochastic approach is indispensable for single-molecule or small-copy contexts.

\section{Conclusion}
Deterministic and stochastic models each shed light on enzyme kinetics from different angles. The deterministic Michaelis--Menten equation (1913), derived via quasi-steady-state arguments, remains highly successful at describing average reaction rates in bulk or large-volume conditions. In small-volume or single-molecule contexts, the stochastic viewpoint captures random fluctuations in enzyme-substrate binding and product formation events such as those in (English, B.P. et al, 2006). From the point of view of Gillespie simulation of the stochastic modeling via the chemical master equation (Gillespie, 2007), we were able to confirm most of the deterministic results under scale, building a bridge between the stochastic and deterministic worlds where the stochastic serves as the underbody of knowledge that serves as a general foundation for enzyme kinetics. For further reading, in a more pure mathematical case, we can look at formalisms like system-size expansions, chemical Langevin equation analysis, and moment-closure approximations to justify the ideas behind simulation. 

Additionally, single-enzyme turnover studies like that of confirm that the MM equation often holds on average (Grimma, 2009). However, the actual distribution of waiting times and product counts can only be captured by stochastic modeling. Noise is not simple and always negligible. It is a technical complication with many resulting mechanistic details and implications. In cellular settings, noise has consequences for reliability and metabolic control, which highlights the relevance of stochastic models in quantitative biology. From a practical standpoint for modeling, it becomes a simple decision: if the system has large molecule counts or if only average rates are needed, the deterministic approach suffices and is computationally efficient. If single-enzyme data, low copy numbers, or distributional properties matter, a stochastic approach becomes necessary. Hybrid or approximate methods (like those mentioned in Wilkinson, 2011) can also be employed to manage complexity. In this way, the synergy of deterministic and stochastic frameworks provides a richer understanding of enzyme kinetics, spanning from classical biochemistry to modern single-molecule and systems biology contexts.

\newpage

\section*{References:}
\begin{itemize}
\item Atkins, William M. "Michaelis-Menten Kinetics and Briggs-Haldane Kinetics." University of Washington, https://depts.washington.edu/wmatkins/kinetics/michaelis-menten.html.
\item Briggs, G. E., \& Haldane, J. B. S. (1925). A note on the kinetics of enzyme action. \emph{Biochem. J.}, 19(2), 338--339.
\item English, B. P., et al. (2006). Ever-fluctuating single enzyme molecules: Michaelis--Menten equation revisited. \emph{Nat. Chem. Biol.}, 2(2), 87--94.
\item Gillespie, D. T. (1977). Exact stochastic simulation of coupled chemical reactions. \emph{J. Phys. Chem.}, 81(25), 2340--2361.
\item Gillespie, D. T. (2000). The chemical Langevin equation. \emph{J. Chem. Phys.}, 113(1), 297--306.
\item Grima, R. (2009). Investigating the robustness of the classical enzyme kinetic equations in small intracellular compartments. \emph{BMC Systems Biology}, 3, 101.
\item Haas, Kathryn. "Michaelis-Menten Kinetics." Chemistry LibreTexts, 2 Mar. 2024, https://chem.libretexts.org/Bookshelves/Biological\_Chemistry/Supplemental\_Modules\_(Biological\_Chemistry)/Enzymes/Enzymatic\_Kinetics/Michaelis-Menten\_Kinetics.
\item Michaelis, L., \& Menten, M. (1913). Die Kinetik der Invertinwirkung. \emph{Biochem. Z.}, 49, 333--369.
\textbf{\item van Kampen, N. G. (2007). \emph{Stochastic Processes in Physics and Chemistry}, 3rd ed. Amsterdam: Elsevier.
\item Wilkinson, D. J. (2011). \emph{Stochastic Modelling for Systems Biology}. 2nd ed. CRC Press.}

\end{itemize}

\begin{description}
    \item[\textbf{* Bolded items were heavily used in this paper.}]
\end{description}

\newpage

\appendix
\section{Code for ODEs Methods and Monte-Carlo Simulation}

This appendix provides the full Python source code used in the project for deterministic and stochastic enzyme kinetics modeling.

\subsection{Deterministic vs. Stochastic Enzyme Kinetics}

\begin{lstlisting}[language=Python, caption={Deterministic vs. Stochastic Enzyme Kinetics Code}]
import numpy as np
import matplotlib.pyplot as plt
from scipy.integrate import odeint

# Deterministic ODE method
def enzyme_odes(y, t, k1, k_1, k2, E0):
    E, S, ES, P = y
    dEdt  = -k1*E*S + k_1*ES + k2*ES
    dSdt  = -k1*E*S + k_1*ES
    dESdt =  k1*E*S - (k_1 + k2)*ES
    dPdt  =  k2*ES
    return [dEdt, dSdt, dESdt, dPdt]

def gillespie_enzyme(k1, k_1, k2, E0, S0, tmax):
    t = 0.0
    nE  = E0
    nS  = S0
    nES = 0
    nP  = 0
    times, E_vals, S_vals, ES_vals, P_vals = [t], [nE], [nS], [nES], [nP]

    while t < tmax:
        a1 = k1 * nE * nS   
        a2 = k_1 * nES      
        a3 = k2 * nES       
        a0 = a1 + a2 + a3
        if a0 <= 0: break

        tau = -np.log(np.random.rand()) / a0
        t += tau
        r = np.random.rand()
        if r < a1/a0:
            nE, nS, nES = nE - 1, nS - 1, nES + 1
        elif r < (a1 + a2)/a0:
            nE, nS, nES = nE + 1, nS + 1, nES - 1
        else:
            nE, nP, nES = nE + 1, nP + 1, nES - 1

        times.append(t)
        E_vals.append(nE)
        S_vals.append(nS)
        ES_vals.append(nES)
        P_vals.append(nP)
        if nS == 0 and nES == 0: break

    return np.array(times), np.array(E_vals), np.array(S_vals), np.array(ES_vals), np.array(P_vals)

# Parameters
k1, k_1, k2 = 0.1, 0.05, 0.05
E0, S0 = 10, 100
y0 = [E0, S0, 0, 0]
tmax = 200
t_points = np.linspace(0, tmax, 2001)
sol = odeint(enzyme_odes, y0, t_points, args=(k1, k_1, k2, E0))
E_det, S_det, ES_det, P_det = sol[:,0], sol[:,1], sol[:,2], sol[:,3]

# Stochastic Simulations
N_runs = 500  
t_grid = np.linspace(0, tmax, 2001)
S_runs, P_runs = np.zeros((N_runs, len(t_grid))), np.zeros((N_runs, len(t_grid)))

for i in range(N_runs):
    t_arr, E_arr, S_arr, ES_arr, P_arr = gillespie_enzyme(k1, k_1, k2, E0, S0, tmax)
    S_runs[i, :] = np.interp(t_grid, t_arr, S_arr)
    P_runs[i, :] = np.interp(t_grid, t_arr, P_arr)

S_mean, P_mean = np.mean(S_runs, axis=0), np.mean(P_runs, axis=0)
S_std, P_std = np.std(S_runs, axis=0), np.std(P_runs, axis=0)

# Plot
fig, ax = plt.subplots(figsize=(7,5))
ax.set_title("Deterministic vs. Stochastic Enzyme Kinetics")
ax.plot(t_points, S_det, label="Deterministic [S]", lw=2)
ax.plot(t_points, P_det, label="Deterministic [P]", lw=2)
ax.plot(t_grid, S_mean, '--', label="Stochastic avg S", lw=2)
ax.plot(t_grid, P_mean, '--', label="Stochastic avg P", lw=2)
ax.fill_between(t_grid, S_mean-S_std, S_mean+S_std, alpha=0.2)
ax.fill_between(t_grid, P_mean-P_std, P_mean+P_std, alpha=0.2)
ax.set_xlabel("Time")
ax.set_ylabel("Molecule Count")
ax.set_xlim(0, tmax)
ax.legend(loc="best")
plt.tight_layout()
plt.show()
\end{lstlisting}

\subsection{Positive Feedback Model}

\begin{lstlisting}[language=Python, caption={Stochastic Positive Feedback Model}]
def gillespie_positive_feedback(alpha0, alpha, K, n, gamma, X0=10, tmax=2000):
    t, X = 0.0, X0
    t_arr, X_arr = [t], [X]

    while t < tmax:
        prod_rate = alpha0 + alpha * (X**n / (K**n + X**n))  
        degr_rate = gamma * X                                
        a0 = prod_rate + degr_rate
        if a0 <= 0: break

        tau = -np.log(np.random.rand()) / a0
        t += tau
        r = np.random.rand()
        X += 1 if r < prod_rate / a0 else -1 if X > 0 else 0

        t_arr.append(t)
        X_arr.append(X)

    return np.array(t_arr), np.array(X_arr)

# Example Parameters
alpha0, alpha, K, n, gamma, X0, tmax = 5.0, 150.0, 20.0, 4, 1.0, 10, 2000
t_arr, X_arr = gillespie_positive_feedback(alpha0, alpha, K, n, gamma, X0, tmax)

# Plot
plt.figure(figsize=(8,4))
plt.plot(t_arr, X_arr, lw=1.5, color='blue')
plt.title("Stochastic Positive-Feedback Model: Random Switching")
plt.xlabel("Time")
plt.ylabel("X (copy number)")
plt.grid(True)
plt.tight_layout()
plt.show()
\end{lstlisting}

\end{document}